%% file: main.tex
\documentclass[10pt, journal, twocolumn]{IEEEtran}
\usepackage{citesort}
\usepackage{color}
\usepackage[latin9]{inputenc}
\usepackage{amsthm}
\usepackage{amsmath}
\usepackage{amsthm}
\usepackage{amssymb}
\usepackage{graphicx}
\usepackage{lipsum}
\usepackage{url}
\usepackage{algorithmic}
\usepackage{algorithm}
\usepackage{mathrsfs}
\usepackage{wrapfig}
\graphicspath{{./}{./Figures/}}
\renewcommand{\figurename}{Fig.}
\usepackage{subfigure}

\input{commands.tex}

\begin{document}

\title{Unleashing the Power of Participatory IoT with Blockchains for Increased Safety and Situation Awareness of Smart Cities}

\author{Bechir Hamdaoui, Mohamed Alkalbani, Taieb Znati$^{\dag}$, Ammar Rayes$^{\ddag}$
\thanks{An IEEE-formatted version of this article is published in IEEE Network. Personal use of this material is permitted. Permission from IEEE must be obtained for all other uses, in any current or future media, including reprinting/republishing this material for advertising or promotional purposes, creating new collective works, for resale or redistribution to servers or lists, or reuse of any copyrighted component of this work in other works.}
~\\
\small Oregon State University, \small Corvallis,
\small \{hamdaoui,alkalbmo\}@oregonstate.edu~\\
$^{\dag}$ \small University of Pittsburgh, znati@cs.pitt.edu~\\
$^{\ddag}$ \small Cisco Systems, rayes@cisco.com
}

\maketitle
\begin{abstract}
\input{Abstract.tex}
\end{abstract}

\section{Introduction}
\label{sec:introduction}
\input{Introduction.tex}

\section{Virtualizable Cloud-of-Things (CoT) Infrastructure}
\label{sec:architecture}
\input{architecture.tex}

\section{Participatory Networks-on-Demand Instantiation}
\label{sec:blockchain}
\input{blockchain.tex}

\section{Edge Cloud Offloading}
\label{sec:offloading}
\input{offloading.tex}

\section{Open Research Challenges}
\label{sec:challenges}
\input{challenges.tex}


\section{Conclusion}
\label{sec:conc}
\input{conclusion.tex}

\section{Acknowledgement}
This work was supported in part by the US National Science Foundation (Award CNS-1162296) and Cisco Systems.

\bibliographystyle{IEEEtran}
\bibliography{References-magazine}

\end{document}

%% file: commands.tex
\newcommand{\comment}[1]{ }

\newcounter{tc}

\newcommand{\myitemizebegin}{\begin{list}{$\bullet$}
{
 \setlength{\leftmargin}{0.4cm}
 \setlength{\parsep}{0.0cm}
 \setlength{\itemsep}{0.05cm}
 \setlength{\topsep}{0.0cm}
}}
\newcommand{\myitemizeend}{\end{list}}

\newcommand{\myitemizebeginless}{\begin{list}{$\bullet$}
{
 \setlength{\leftmargin}{0.7cm}
 \setlength{\parsep}{0.0cm}
 \setlength{\itemsep}{0.05cm}
 \setlength{\topsep}{0.0cm}
}}
\newcommand{\myitemizeendless}{\end{list}}

\newcommand{\ac}{{\tt autCloud}}
\newcommand{\ec}{{\tt coreCloud}}
\newcommand{\edc}{{\tt edgeCloud}}

\newcommand{\coti}{$\tt {CoT~Infrastructure}$}
\newcommand{\IoTagent}{{$\tt IoT~Device~Broker$}}
\newcommand{\VNEagent}{{$\tt NoD~Service~Agent$}}
\newcommand{\svs}{{SVS}}

\newcommand{\NoD}{{NoD}}

\newcommand{\arrivalpara}{{p}}
\newcommand{\durationpara}{{q}}



%% file: Abstract.tex
IoT emerges as an unprecedented paradigm with great potential for changing how people interact, think and live. It is making existing Internet services feasible in ways that were previously impossible, as well as paving the way for new situation-awareness applications suitable for smart cities, such as realtime video surveillance, traffic control, and emergency management. These applications will typically rely on large numbers of IoT devices to collect and collaboratively process streamed data to enable real-time decision making.
%
In this paper, we introduce the concept of {\em Semantic Virtual Space (\svs)}, an abstraction for virtualized cloud-enabled IoT infrastructure that is commensurate with the goals and needs of these emerging smart city applications, and propose and discuss scalable architectures and mechanisms that enable and automate the deployment and management of multiple \svs~instances on top of the cloud-enabled IoT infrastructure. 

%% file: Introduction.tex
IoT has been broadly defined as an ecosystem of smart objects that interact autonomously with each other, fundamentally altering how humans interact with the natural world.
One of its enormous impacts lies in making existing smart city services feasible in ways that were previously impossible, as well as in paving the way for a wide range of new smart city applications, ranging from video surveillance and traffic control to emergency management and precision health.
These applications typically involve monitoring {\em space} and {\em objects} and, in some cases, the interactions between them, and do so by relying on large numbers of IoT devices with sustained one-to-one---and possibly one-to-many---device connectivity to collect and collaboratively process streamed data to enable real-time decision making.

\subsection{Unleashing the Power of Participatory IoT}
In this work, our vision of IoT transcends a mere object-centric view, and considers IoT as a {\em distributed and Internet-accessible infrastructure} that seamlessly integrates the physical and virtual worlds with capabilities far exceeding the computational intelligence, functionality and reliability of today's systems. This IoT infrastructure may serve multiple entities and groups (private or public) of people within a city, where the interest of each entity in collecting information derives directly from the mission of the entity itself.
The factors that impact interactions with the physical infrastructure include heterogeneity among participating organizations and groups, asymmetry in information processes among the groups, and asynchronous dissemination of critical information to participating groups. Embedding {\bf''semantic views''} into IoT to support group interests and services would require intelligent management of the physical and computing infrastructure in order to personalize and adapt to situational and environmental conditions, as required by the supported application.

\subsection{The Concept of Semantic Virtual Space (\svs)}
\begin{figure}
\center{
\includegraphics[width=\columnwidth]{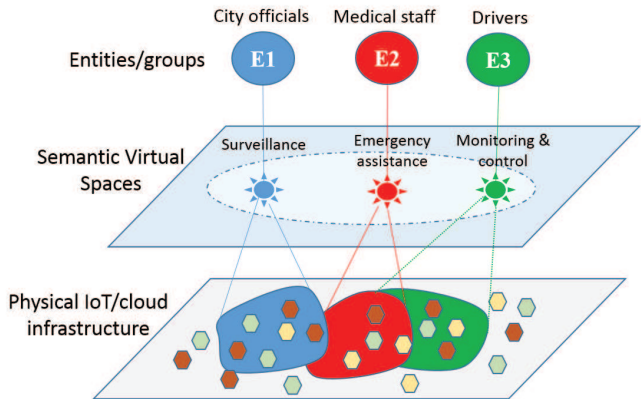}
\caption{{\small Semantic virtual space}}
\label{fig:views}}
\end{figure}

To address the above challenge, we propose the concept of {\em \bf Semantic Virtual Space (\svs)}, an abstraction for virtualized cloud-enabled IoT infrastructure commensurate with the goals and needs of the associated organization and underlying application.
To illustrate this concept, consider a large event like the Soccer World Cup or the Olympic Games taking place in some major city. An event of this scale usually attracts lots and lots of people and is also usually organized for a fixed period of time (e.g., a month) during which, the city is often faced with some major challenges, including resolving traffic congestion, ensuring coordinated and easy access to attractions (e.g., parking, restaurants, hotels, etc.), providing realtime surveillance for people's and city's safety, being well-prepared for emergency relief operations (e.g., accident, fire, etc.), and monitoring and controlling health-related matters (e.g., pollution, disease epidemics, etc.). Our vision is that for such a large-scale event, the city-wide IoT infrastructure can be leveraged to address such challenges. It can, for example, be used to enable and support applications serving three different city entities with different missions: (i) alert police and city officials about security threats that can be identified through realtime video surveillance; (ii) guide medical staff (e.g., ambulance) efficiently through traffic to offer its aid as quickly as possible during emergency relief operations; and (iii) assist and guide visitors to easily find their points of interest (e.g., hotels, restaurants, etc.).
Each of these three missions constitutes a different \svs, and may involve only some components of the physical/IoT infrastructure.
Fig.~\ref{fig:views} depicts the multi-layered architecture supporting these three \svs s, each designed to capture the interest and mission of each of the three different city entities.

An instantiation of an \svs~to support a situation-awareness application consists then of a {\em dynamic, on-demand} logical grouping of a set of geographically distributed participatory IoT devices, created to process, filter and fuse collected data into accurate and actionable information for realtime decision making, as required by the underlying application~\cite{blockchains-iot-hamdaoui}.
Throughout, we will be referring to each \svs~instantiation as a {\bf participatory IoT network-on-demand (\NoD) instance}.

In this work, we propose scalable architecture and mechanisms that allow the instantiation, deployment and management of multiple participatory \NoD~instances to enable and automate a wide range of situation-awareness and safety smart city applications.
These targeted applications share key characteristics and requirements. Firstly, they require {\em interactive} execution among, and involvement of, the devices participating in the \NoD~instance. For example, for the realtime video surveillance case, measurements made by individual cameras can be noisy, and therefore, collective measurements are needed to refine the estimates and avoid false detections. Besides, when the surveillance system is being used to track moving objects, multiple different field views may be needed to be able to make reasonably accurate decisions.
Secondly, {\em realtime} extraction of actionable knowledge is needed to be able to take timely actions. For example, in the case of an emergency management instance, it is important that the security officials and medical staff be informed immediately of what happens, what to do, and where to go, so that necessary actions are taken timely.
Thirdly, they may only be needed {\em temporarily}, typically days or weeks, as for the case of sports and concert events, thus calling for elastic and virtualizable resource provisioning solutions to allow for resource scaling.
These aforementioned requirements signal then a paradigm shift from the traditional {\bf `collect data now and analyze it later'} approach to the {\bf `collect, analyze and decide on the fly'} approach, and our proposed framework distinguishes itself by leveraging key emerging technologies like edge cloud computing, IoT and blockchains to allow and ease such a paradigm shift.

This paper is organized as follows. We first present the proposed architecture in Section~\ref{sec:architecture}. Our architecture couples IoT device potentials with cloud computing capabilities~\cite{chun2011clonecloud} to enable our envisioned situation-awareness smart city applications, and throughout, we will refer to this architecture as {$\tt CoT~(Cloud~of~Things)~Infrastructure$}. We then present the proposed blockchain-enabled distributed mechanisms in Section~\ref{sec:blockchain}, and the edge cloud offloading techniques in Section~\ref{sec:offloading}. Finally, we highlight key open research challenges in Section~\ref{sec:challenges}, and conclude the paper in Section~\ref{sec:conc}.

%% file: architecture.tex
\begin{figure}
\center{
\includegraphics[width=\columnwidth]{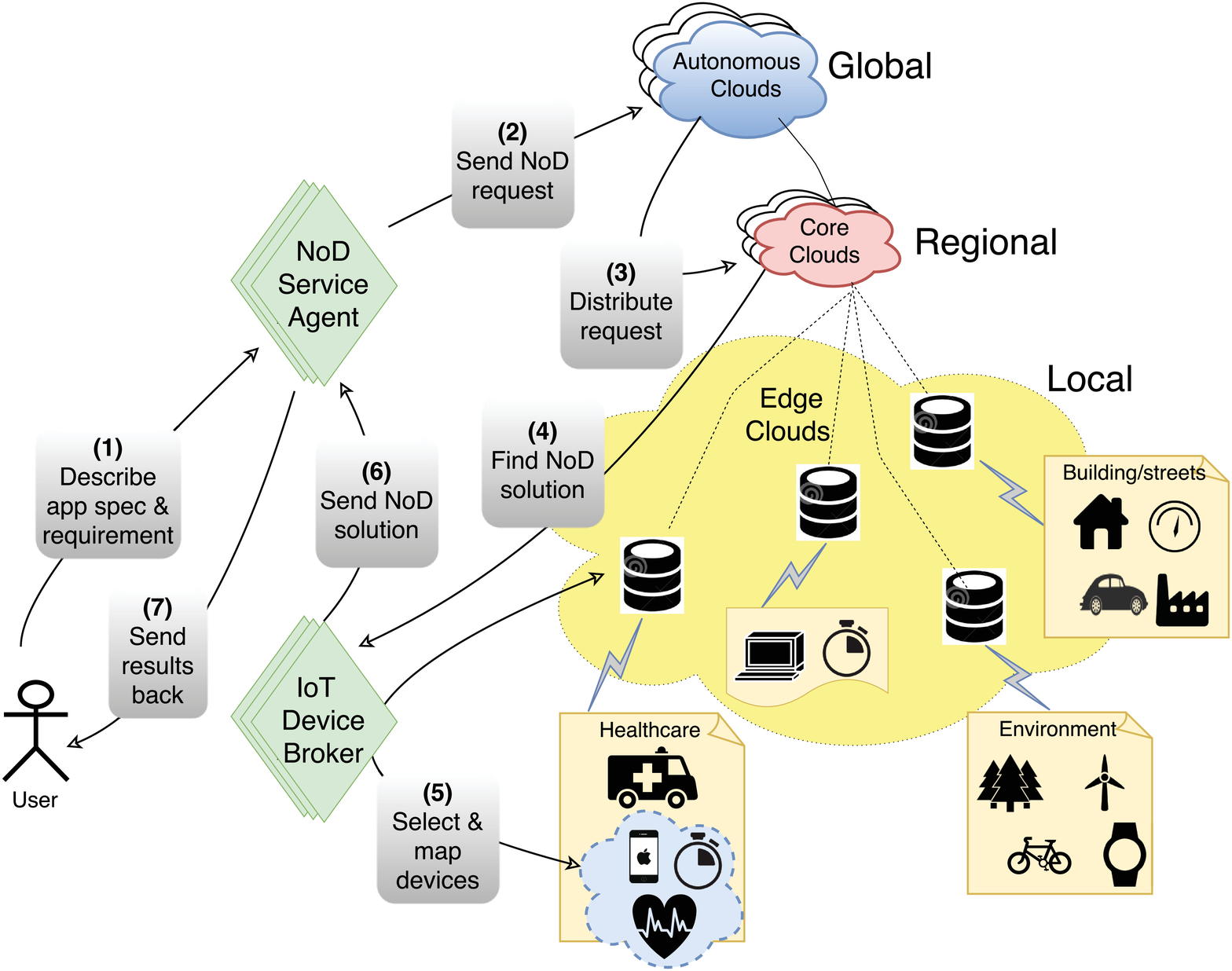}
\caption{{\small \coti}}
\label{fig:overview}}
\end{figure}
\subsection{Scalable Architecture for Elastic and Fast \NoD~Instantiation}
Our envisioned architecture is cloud-enabled IoT infrastructure, referred to as \coti. It consists of 4 tiers (Fig.~\ref{fig:overview}). The top tier, Tier 1, contains the different autonomous clouds (\ac s), which are the logical entities that own conventional cloud platforms (e.g., Amazon clouds) and provide interfaces for user access.
Each \ac~typically covers multiple regions in the world, each constituted of several (core) clouds (\ec s), where, in most cases, a \ec~is nothing but a datacenter. In this architecture, the set of all \ec s forms Tier 2. Tier 3 constitutes the set of all edge clouds (\edc s), which are essentially small-scale datacenters deployed at the network edge in a city to bring data and computing closer to the IoT devices.
For example, LinkNYC\footnote{\url{https://en.wikipedia.org/wiki/LinkNYC}}~\cite{hassan-gc:16,sinky2018responsive}, an infrastructure project announced in 2014 and became operational in 2016,
replaces thousands of payphones with kiosk-like structures, called Links, to offer fast, free Wi-Fi access to everyone in New York City.
When equipped with appropriate computing and storage resources, these Links can be viewed in the case of New York City as \edc s. Finally, Tier 4 represents the IoT devices where each device is to be associated with one (or more) \edc(s) for connectivity, accountability, and service purposes.

\subsection{Service Provider Entities: Functionalities and Interactions}
We envision that the architecture proposed above will trigger the emergence of new service/bunisness provider entities. Here, we introduce two: \IoTagent s~and \VNEagent s.
\IoTagent s are perceived as business entities that serve as brokers for the {\em participatory} IoT devices, and can, for example, be responsible for handling the registration and authentication of IoT devices (e.g., obtaining their resource capacity, location, mobility information, etc.), defining and managing payments and other associated logistics, assigning registered IoT devices to appropriate \edc s, and publishing and making this information available to other entities. \VNEagent s, on the other hand, are high-layer service providers that serve as the liaison between \NoD~clients, cloud platform providers, and \IoTagent s. They can, for instance, be responsible for receiving \NoD~requests from the different interest groups (e.g. law enforcement officials, interested in tracking a suspected criminal in some area, can issue a request describing the requirements and specifications of their surveillance application to the \VNEagent) (Step 1 in Fig.~\ref{fig:overview}).
In turn, these agents translate these requirements and specifications into \NoD~requests,
and send them to the different \ac s (Step 2) so that \NoD~requests are disseminated to the \ec s covering the area of interest (Step 3), as specified in the request, triggering then the execution of the \NoD~instantiation mechanisms (discussed later). When a \NoD~request needs data from devices registered with different \ac~entities, \VNEagent~can facilitate this task by federating across the different \ac s. \ec s in collaboration with \IoTagent s then run the instantiation mechanisms to discover and locate physical network resources (Step 4) and perform the resource mapping task (Step 5).
The \NoD~solution is then sent back to
the \VNEagent~(Step 6), which monitors the created virtual \NoD~instance and sends its information (e.g., configuration, control, etc.) back to the client (e.g. law enforcement officials) (step 7).

%% file: blockchain.tex
We now present our mechanisms proposed to manage participatory IoT devices and map \NoD~instances on top of these devices. We want to mention that the focus of this work is on the \ec s, \edc s and IoT devices layers of the system; that is, the city-level architectural components.
Our mechanisms leverage blockchain technology to allow: the registration, discovery and management of IoT devices wanting to participate in \NoD~instances, enable the mapping of requested \NoD~instances onto the registered IoT devices, ensure service delivery and integrity of committed IoT devices, and reward and secure payments to the IoT devices participated in the \NoD~instances.

Although blockchain technology~\cite{nakamoto2008bitcoin,swan2015blockchain} is conventionally used for cryptocurrency, due to its distributed nature and great potential in simplifying recordkeeping, it has been attracting many other applications (e.g., voting, vehicle registrations, IoT applications, etc.). In this work, we leverage it to design distributed mechanisms for scalable and fast \NoD~instantiations. Adopting blockchain features in \NoD~instantiation mechanisms is not, however, straightforward, and presents new challenges, pertaining to IoT and arising from the following facts and features of the system at hand: (i) IoT devices have limited storage and computation resources, (ii) the situation awareness applications supported by \NoD~instances are delay sensitive, and (iii) the bandwidths available for these IoT devices could be limited (e.g., wireless connections). The design approaches we present in this paper aim to address these challenges.

\subsection{Blockchain-Enabled \NoD~Instantiation Mechanisms}
We consider a city-wide \coti~constituted of many IoT devices spread all over the city, and a set of \edc s also deployed across the city to provide Internet connectivity and resource offloading to the IoT devices. An IoT device interested in making side income by participating in \NoD~instances needs to advertise, upon joining the network, its device characteristics (e.g., resource type/capacity, availability, bounty, etc.) to the devices in the network including (some of) the \IoTagent s  overseeing and offering service in that city.
As mentioned earlier, in our architecture, \IoTagent s are responsible for receiving and handling the \NoD~requests, and serve as the liaisons between the requests and the registered IoT devices by enabling mechanisms and protocols that allow the creation and management of such \NoD~instances.
The proposed blockchain-based mechanisms consist of two major components, each playing an essential role towards achieving our ultimate goal of enabling scalable and fast \NoD~deployments.

\subsubsection{Registration, discovery and mapping component}
(i) Allows participatory IoT devices to join, authenticate and register themselves to the network. (ii) Enables the discovery of IoT devices satisfying the requirements of the \NoD~requests, based on devices' reputations, prices, capacity, availability, etc. And (iii) enables the mapping of the \NoD~requests on top of the discovered IoT devices to create the \NoD~instances. This component has two phases:
\myitemizebegin
\item {\em Device authentication and registration phase:} Each participatory IoT device is required to register by broadcasting its device characteristics to all other devices in the city-wide network. Device characteristics include information such as device ID, device wallet ID, public key, resource type, resource capacity, availability, reputation score, bounty, etc. Upon joining the network, device reputation and wallet values will be set to zero, which will be updated, as discussed later, as the device starts participating in \NoD~instances. This information will be digitally signed (via public/private key) before broadcast for authentication and integrity purposes, and will be added to the blockchain by miners. It will also be used later to verify and confirm whether a device meets the requirements of \NoD~requests and thus can be mapped to any of the requests to serve as a participatory device.

\item {\em Resource discovery and mapping phase:} All \NoD~requests will be received and handled by the \IoTagent s, and the brokers will serve as the liaison between the devices and requests. For each received \NoD~request, to create the \NoD~instance, the broker will disseminate the request information to a set of (one or more) devices covering different regions of the city of interest to the \NoD~request.
    An \NoD~request is modelled with a tuple $G=(N,R,L,T,C,S)$, with $N$ being the number of needed devices, $R$ being the type of needed resources, $L$ being the locations of devices, $T$ the time during which these resources will be needed, $C$ the cost/bounty the broker is willing to pay a participatory device for its service, and $S$ the minimum reputation score a device needs to have to be able to participate. The request will be circulated among the devices, and as it goes through them, a device meeting the requirements of the request can choose to join the \NoD~instance. And if it does, it updates the request accordingly, and forwards it to other devices. By the end of this phase, the devices participating on the requested \NoD~will all be selected, and the \NoD~instance will be created. The broker will assign a unique ID for each created \NoD~instance for accountability and manageability. Once created, the \NoD~instance can start running the underlying application as requested, where devices will be using their resources to perform their assigned tasks, and possibly be communicating with one another as dictated by the supported application. Network traffic flow configuration and control, which can be managed via SDN, is beyond the scope of this work.
\myitemizeend

\subsubsection{Verification, payment and accountability component}
(i) Ensures that the IoT devices committed to \NoD~instances perform their tasks as agreed upon. (ii) Provides backup plans for those devices that fail to deliver their service. (iii) Secures payment operations and fund transfers from consumers to participatory devices upon completion of their assigned tasks. And (iv) employs a trust mechanism that allows devices' reputations to be built-up and updated based on their successful completion of assigned tasks.
These capabilities will be enabled through the three following phases:

\myitemizebegin
\item {\em Service delivery monitoring phase:}
As \NoD~requests are disseminated through the different participatory devices during the discovery and mapping phase described above, a set of devices will be selected to serve as monitors whose job is to probe the committed IoT devices periodically to make sure that they are still up to perform their agreed upon tasks.
Whenever a monitor notices that a committed device is not responding to its probes or observes malicious activity inferring that the device is not performing its assigned task, the monitor raises a flag and informs all other devices. This can, for example, be used to trigger a replacement of the failing device, and to rate devices for their offered service quality to update their reputation and trust levels, as discussed next.

\item {\em Building trust and reputation phase:}
Unlike Bitcoin, which manages cryptocurrency transfers among users and hence verification can easily be done by looking at the transfer's wallet/balance, verifying that committed IoT devices performed their tasks as agreed upon is not a task that can easily and accurately be checked by monitoring devices.
This is similar to online shopping (e.g., eBay) and car transportation markets (e.g., Uber), where service delivery can be confirmed only after the goods/services are received/delivered.
Therefore, like these systems, we rely on reputation-based schemes to score and select participatory devices. Here, each IoT device participated in an \NoD~instance receives a score for its delivered service quality, which is then used to build its reputation for future participation.
In addition to \IoTagent s, monitors, as well as devices participated in the same \NoD~instance, rate devices too, and a voting mechanism is used to decide on the final rating.
Once the rating is decided on, the newly updated reputation score is broadcast to all devices, and is included in the new block to be added to the blockchain.

\item {\em Service delivery verification and payment phase:}
Now it needs to be decided whether a device performed its service as it should so funds can be transferred to it. We also rely on voting approaches to make such decisions. Once a fund transfer is voted on, this transfer transaction is broadcast to all devices and is added to the blockchain too.
\myitemizeend

\subsection{Performance Results}
{We consider a time-slotted system where at each time slot, a new \NoD~request, with service duration (in number of time slots) following a Bernoulli process with parameter $\durationpara$, arrives according to Bernoulli random variable with parameter $\arrivalpara$.
We define the {\em network load} as $\arrivalpara/\durationpara$, which essentially represents the average number of \NoD~requests that would have been present in the network at a time slot had all arrived requests been accepted. Or said differently, the {\em average number of \NoD~requests that are actually present in the network} is the {\em network load} multiplied by the corresponding {\em acceptance rate} of arrived requests.
In this experiment, we set $\arrivalpara$ and $\durationpara$ in such a way that the network load varies between 0.2 and 0.6. Also, the number of requested devices $N$ is set to 10, the locations $L$ of the 10 requested nodes are selected randomly within the city, the request bounty $C$ is selected uniformly between 100 and 1000, the size of mining period (during which on block is added to the blockchains) is set to 3 time slots, and the number of monitors is set to 3.}

\subsubsection{Device discovery and \NoD~request mapping}
\begin{figure}
\centerline{
    \subfigure[\scriptsize Acceptance rate.]
    {\includegraphics[width=\columnwidth]{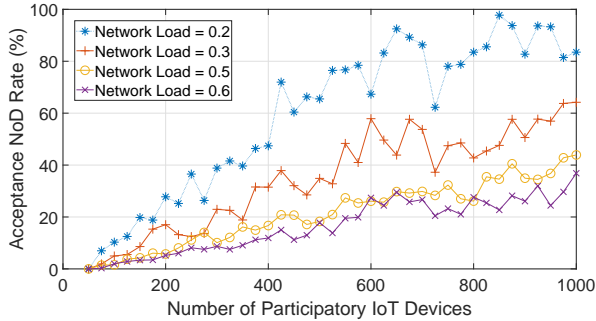}
    \label{subfig:acceptance}}}
\centerline{
    \subfigure[\scriptsize Number of visited devices.]
    {\includegraphics[width=\columnwidth]{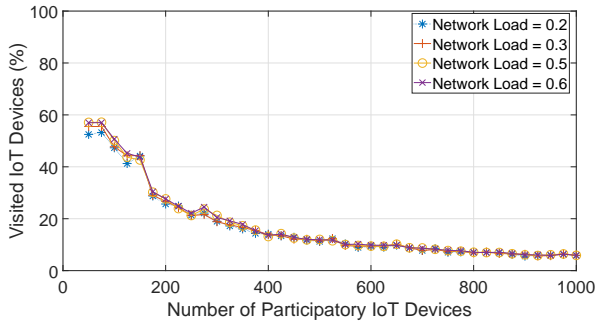}
    \label{subfig:visited}}}
    %
\caption{{{\small Mapping performance results under reliable networks (no node failure) with request size = 10.}}}
\label{fig:bc-perf}
\end{figure}

We first study the impact of the network load on: (1) {the acceptance rate} of \NoD~requests and (2) the {average number of visited devices} before the \NoD~request is successfully mapped.
Fig.~\ref{subfig:acceptance} shows the acceptance rate under different network loads. As expected, observe that the acceptance rate increases with the network size and decreases with the network load.
{This is because as the network size increases, the likelihood of finding nodes that can be mapped onto the requests increases, resulting in higher acceptance rates. But as the network load increases, more nodes in the network become committed to requests, making it harder to find devices that can meet the new requests' requirements, thus resulting in lower acceptance rates.}
Fig.~\ref{subfig:visited} shows that the number of IoT devices visited before a mapping is found decreases as the number of participatory devices increases, since again the likelihood of finding devices that meet the request's requirement increases with the size of the network. However, our results show that such a tendency is not as dependent on the network load as in the case of the acceptance rate metric.
{This is because both the already committed and non-committed devices have to be visited to check for their availability prior to accepting the request, and hence, a higher percentage of committed nodes does not impact the number of nodes that need to be visited to fulfill a request.}

\begin{figure}
  \centering
  \includegraphics[width=\columnwidth]{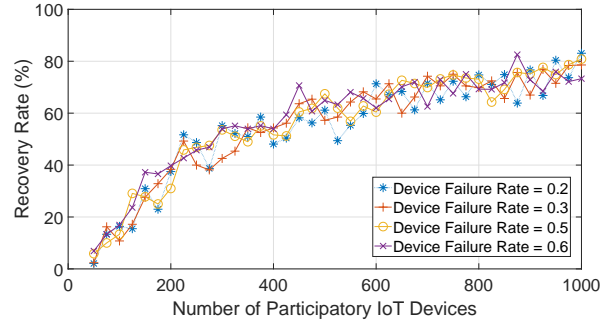}
  \caption{{{\small Mapping recovery rate under different device failure rates for network load = 0.5.}}}
  \label{fig:perf-failure}
\end{figure}

\subsubsection{Robustness to node failures}
We also consider the case when devices could fail during and/or after the mapping of requests, and propose a failure-recovery mechanism that incorporates (1) monitoring and detection capability, which allows to track and check whether committed devices are still up to the assigned task, and (2) re-mapping capability, which allows to find a quick replacement to failed devices. To have a sense of how our recovery mechanism performs, we show in Fig.~\ref{fig:perf-failure} the recovery rate of the proposed mechanism by measuring the ratio of the number of successfully recovered requests to the total number of failed requests.
{First, note that as the number of IoT devices in the system increases, the recovery rate increases, regardless of the device failure rate. This is because the higher the number of nodes, the greater the likelihood of finding nodes that satisfy the failed nodes' requirements, thus increasing the recovery rate. Note that the recovery rate could reach up to 80\% for reasonable sizes of networks (e.g., 1000).
Second, note that the node failure rate has little effect on the recovery rate. This is because as the device failure rate increases, the recovery mechanism can still recover from failures that happen to different nodes in the network. Since the load is constant, the likelihood of finding a node that satisfies the failed network requirement is the same.}

\subsubsection{Blockchains robustness to the 51\% attack}
\begin{figure}
  \centering
  \includegraphics[width=\columnwidth]{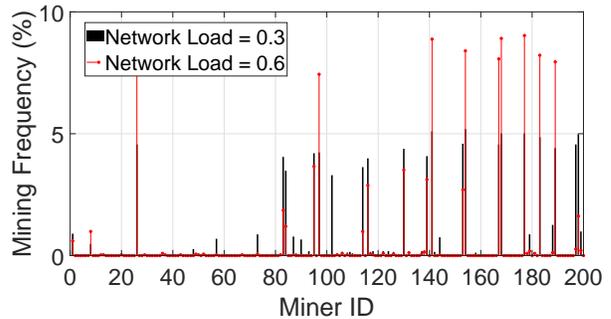}
  \caption{{\small Miner selection distribution: node failure rate = 0.2 and number of miners = 200.}}
  \label{fig:perf-mining}
\end{figure}

We now provide some results assessing how robust our blockchains-enabled mechanism is to the 51\% attack~\cite{bentov2014proof}. For this, we measure and plot in Fig.~\ref{fig:perf-mining} the mining frequency (the number of times a miner has been selected as a miner divided by the total number of miner selections or mining periods) of each miner under two different network loads. The network in this experiment contains 200 miners. The figure shows that no miner has been selected more than 9\%, and no miner has been selected overwhelmingly more than the other miners. This demonstrates that our blockchains-enabled mechanism is robust to the 51\% attack.

%% file: offloading.tex
In addition to enabling resource provisioning elasticity that allows to scale up and down resources as needed, edge cloud offloading offers two key benefits~\cite{cuervo2010maui,shi2014cosmos}. {Firstly, it} provides great incentives for IoT device participation, as it exempts them from having to deal with the computation and storage burdens of the supported application, and {secondly}, it improves IoT application responsiveness by reducing end-to-end latency.

\subsection{Device Cloning for Edge Cloud Offloading}
One possible way for enabling edge cloud offloading is to {\em clone} the IoT devices at the network edge through the creation of dedicated virtual machines (VMs)~\cite{chun2011clonecloud,satyanarayanan2009case,chun2009augmented}.
{The concept of cloning wireless devices (IoT, smart phones, and others) at the edge cloud is introduced to essentially mitigate the resource (CPU, power, etc.) limitations of such wireless devices by offloading their task computation/execution to the cloud~\cite{satyanarayanan2009case,chun2009augmented}. Task execution offloading via cloud cloning involves four steps~\cite{chun2009augmented}: (i) a clone of the IoT device is first created and hosted in the closest edge cloud; (ii) the state of the device and its clone is synchronized reactively (when there is change) or periodically; (iii) task is executed (partially or fully) in the clone, automatically or upon request; (iv) execution outcome of the clone is re-integrated back to the primary device.
Edge cloud offloading via device cloning offers thus three key benefits.} First, the clone can itself provide message brokering services, so that other devices participating in the same \NoD~instance can, through their own clones, communicate faster with one another, as their communication will be among and through the cloud clones. Second, cloning reduces the communication and computation burden of those devices that participate in multiple, concurrent \NoD~instances. For example, a camera deployed in a city street can be taking video data to serve three situation-awareness applications concurrently, each supporting a different interest group; e.g., help locate street parking spots, provide video surveillance, and assist emergency personnel during relief operations.
Cloning can be very handy in such scenarios, as it eliminates the need for the device to communicate with the edge multiple times, one for each \NoD~instance. For this, each instance, implemented for example via a process, can just {\em subscribe} to the device clone, allowing it to receive the video content of relevance to it directly from the clone. Third, it exempts the device from any computation and device-to-network communication that may be needed during the running of the \NoD~instantiation mechanisms.

\subsection{Online Clone Migration for Optimal Cloud-Clone Resource Mapping}
Allowing dynamic migration of clones across the different \edc s is important to ensure that resources are allocated efficiently and application requirements are guaranteed to be met at all times.
As a result, few techniques (e.g.,~\cite{wang2015mobiscud}) emerged to allow clone migration so that latency is kept at minimum, where migrations, in these approaches, are triggered mainly based on device mobility.
However, in our envisioned situation awareness IoT applications,
device clones belonging to the same \NoD~instance will have to communicate with one another, as well as with their devices, making their interactions a determining factor for deciding whether and if to move, as opposed to just relying on device mobility. In an effort to address this issue, {\em Flock} \cite{abdelwahab2017flocking,abdelwahab2018clones} is proposed to allow live migration of clones to be triggered based not only on device mobility, but also on inter-clone traffic behaviors and demands as dictated by the underlying application, thereby improving application responsiveness and resource allocation efficiency. Flock imitates the {\em bird flocking behavior}~\cite{flocking-behavior}, controlled by three known rules,
separation (avoid crowding clones), alignment (steer towards average heading), and cohesion (steer towards average position), to allow clones to be migrated autonomously between the different \edc s so that end-to-end latencies are minimized~\cite{abdelwahab2017flocking}.

%% file: challenges.tex
\subsection{Architectural and Functional Design Challenges}

\subsubsection{Architectural entities and their interactions}
With respect to the proposed architecture, there remains a need to identify and clearly spell out the different architectural and service entities, define their roles, functionalities and responsibilities, and specify their interfaces and interactions. For instance, \IoTagent's responsibilities may include managing the registration and the monitoring of IoT devices, a task that can be very challenging due to the heterogeneity as well as the number of IoT devices at hand. For ease of manageability, \IoTagent s may therefore need to work out careful taxonomy of IoT devices that could for e.g. be domain based (health, traffic, etc.), ownership based (participatory, public, enterprise, etc.), or mobility based.

\subsubsection{Unified interactive language}
Due to the complexity of the \coti~at hand, many types of deal-making agents (brokers, negotiators, auctioneers, regulators) will emerge in this system, with each agent having different needs and requirements for its interactions with the other agents. Therefore, ensuring that all the different entities and agents use unified language with common concepts and constructs that eases their interaction and allows them to express their requirements and preferences and to learn to function in such a complex system is crucial to the successful deployment of these \NoD s.

\subsubsection{Intercloud interoperability}
Intercloud interoperability eases data deployment and migration across different clouds for better resource sharing, and provides the flexibility to select, mix, and/or change cloud service providers with minimal input and intervention. It also facilitates adoption of new elements to the clouds, and allows software, protocol, and/or technology reusability across different cloud platforms.
Although intercloud interoperability has already been recognized as an important topic, very little has been done so far to address its challenges. Some open challenges are the definitions and derivations of metrics that quantify and assess whether service providers met their obligated service-level agreements (SLAs), as well as the development of algorithms and tools that can be used to assess such metrics.

\subsubsection{Manageability and control of \NoD~instances}
Significant research has leveraged SDN and NFV to ease network management and control through the creation of network abstractions and APIs.
This led to the development of new technologies and protocols like OpenFlow, which have gained widespread deployment and usage in a variety of controllers and network environments.
Similar efforts have focused on developing application-aware techniques to ensure QoS guarantees, exploiting SDN and NFV in mobile networking and edge computing.
As a result, a number of SDN- and NFV-centric dynamic resource allocation frameworks have been proposed with increasing deployment in network environments and cloud computing infrastructures. Very little work, however, has focused on supporting the deployment and instantiation of participatory \NoD.

\subsection{Blockchain Challenges}
In Bitcoin, miners are selected on a Proof-of-Work (POW) basis by solving computationally-heavy puzzles. Although Bitcoin's POW requirement ensures system robustness (e.g., tackles double spending and 51\% attack problems), it can't be used in our framework, simply because IoT devices are not powerful and our underlying applications are not delay tolerant.
Therefore, new mining approaches suitable for IoT that can ensure system robustness but without incurring heavy computation and long delays need to be investigated.
{For instance, for miner selection, one approach to consider is to allow multiple miners to mine for the same block; for example, IoT devices can all mine on a first-come, first-served basis, and stop mining when and after some number of devices succeed.
Proof-Of-Stake based selection approaches, which do not require devices to solve puzzles but instead rely on devices' stakes in the system to decide on how one can serve as a miner, could be the appropriate mining strategy for such systems, but further research needs to be conducted in this regard.
Another idea to investigate is to allow \IoTagent s to serve as miners too; since only \IoTagent s serve as consumers in our system, the double spending problems will be inherently solved.
Also, unlike in Bitcoin, where different miners succeeding in finding a nonce generate different hashes/blocks, in our case, devised approaches need to allow all miners to generate the same block to ensure consistency among multiple miners.}

\subsection{Edge Cloud Offloading Challenges}

\subsubsection{Device-clone interaction}
To harness the benefits of cloning, questions like how often should each IoT device upload its data to its clone, and which data to upload remain to be answered.
Also, some IoT devices may change their locations, and if so, how should clones be handled in this case? One way is to allow clone migration, which  can handle mobility, in addition to maintaining low latency and high resource utilization. However, there clearly exist tradeoffs between migration cost and performance gain that need to be investigated.

\subsubsection{Live clone migration}
Although live migration approaches have already been proposed, there remains an urgent need for techniques that are suitable for the envisioned IoT applications. Key design requirements that need to be accounted for are: (i) Triggering clone migrations not just via device mobility but also via changes in inter-clone traffic behavior and conditions and clone-to-clone relationships as dictated by the application.
(ii) Enabling distributed migration by relying on local measurements that clones can collect through simple interactions.
(iii) Incorporating inter-clone traffic behavior and demands into the cloud selection mechanism to improve responsiveness.
And (iv) promoting design simplicity by enabling clone migration without requiring changes to existing cloud platform controllers.

%% file: conclusion.tex
This paper proposes distributed architectures and mechanisms that exploit edge cloud computing and blockchains technologies to enable scalable and elastic deployment of participatory IoT networks-on-demand with the goal of supporting situation-awareness and safety applications in smart cities.
{Specifically, it proposes the concept of {\em \bf Semantic Virtual Space (\svs)}, which is an abstraction for a dynamic, cloud-enabled IoT infrastructure that is commensurate with the goals and needs of the supported smart city applications.
\svs~leverages edge cloud technology to help mitigate the resource limitation of IoT devices, and blockchain technology to ease and enable distributed management of participatory IoT devices at scale.
The paper also discusses the vital role edge cloud computing plays when it comes to enabling IoT device offloading and elastic resource provisioning, thereby improving the responsiveness of IoT devices and the applications they support, as well as their incentives for participation.}
The paper finally describes a set of open research challenges, pertaining to enabling participatory IoT networks-on-demand through edge clouds and blockchains. 